\documentclass[showpacs,twocolumn,aps,prl]{revtex4}
\usepackage{amsmath}
\usepackage{amsfonts}
\usepackage{amssymb}
\usepackage{amsthm}
\usepackage{graphicx}

\begin{document}

%----------------------------------------------------------------%
%\title{Theoretical limits on the magnetization switching field
%and switching time for Stoner particles}
\title{Theoretical limit of the minimal magnetization switching
field and the optimal field pulse for Stoner particles}
\author{Z. Z. Sun}
%\email[To whom correspondence should be addressed. Electronic
%address: ]{phszz@ust.hk}
\affiliation{Physics Department, The Hong Kong University of
Science and Technology, Clear Water Bay, Hong Kong SAR, China}
\author{X. R. Wang}
\affiliation{Physics Department, The Hong Kong University of
Science and Technology, Clear Water Bay, Hong Kong SAR, China}
\date{\today}

\begin{abstract}
The theoretical limit of the minimal magnetization switching
field and the optimal field pulse design for uniaxial Stoner
particles are investigated. Two results are obtained.
One is the existence of a theoretical limit of the smallest
magnetic field out of all possible designs. It is shown that
the limit is proportional to the damping constant in the weak
damping regime and approaches the Stoner-Wohlfarth (SW) limit at
large damping. For a realistic damping constant, this limit is
more than ten times smaller than that of so-called precessional
magnetization reversal under a non-collinear static field.
The other is on the optimal field pulse design: If the magnitude
of a magnetic field does not change, but its direction can vary
during a reversal process, there is an optimal design that gives
the shortest switching time. The switching time depends on the
field magnitude, damping constant, and magnetic anisotropy.
However, the optimal pulse shape depends only on the damping
constant.
\end{abstract}
%\keywords{superlattice, SSCO, phase diagram}
% 75.60.Jk Magnetization reversal mechanisms\\
%75.75.+a Magnetic properties of nanostructures \\
%85.70.Ay Magnetic device characterization, design and modeling \\
\maketitle
%----------------------------------------------------------------%
Fabrication\cite{Sun,Qikun} and manipulation\cite{Hillebrands}
of magnetic single-domain nano-particles (also called the
Stoner particles) are of great current interests in
nano-technology and nano-sciences because of their importance
in spintronics. Magnetization reversal, which is about how
to switch a magnetization from one state to another, is an
elementary operation. One important issue is how to switch
a magnetization fast by using a small switching field.
The switching field can be a laser light\cite{Bigot}, or a
spin-polarized electric current\cite{Slon,current}, or
a magnetic field\cite{field,xrw}.
Many reversal schemes\cite{xrw1,bauer} have been proposed
and examined. However, the issue of theoretical limits of the
smallest switching field and the shortest switching time under
all possible schemes are not known yet.
Here we report two theorems on the magnetic-field induced
magnetization reversal for uniaxial Stoner particles.
One is about the theoretical limit of the smallest possible
switching field. The other is about the optimal field pulse
for the shortest switching time when the field magnitude is
given.

\begin{figure}[htbp]
 \begin{center}
\includegraphics[width=8.cm, height=4.cm]{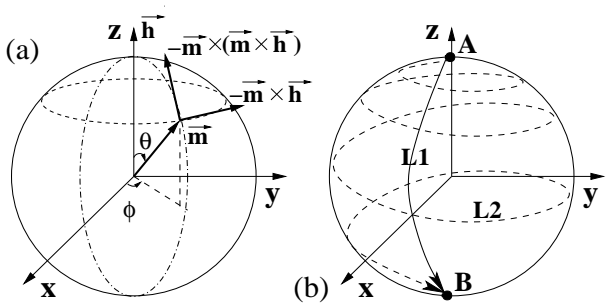}
 \end{center}
\caption{\label{fig1} {\bf a}, Two motions of magnetization
$\vec{m}$ under field $\vec{h}$: $-\vec{m}\times \vec{h}$ and
$-\vec{m} \times (\vec{m}\times \vec{h})$ describe the precession
and dissipation motions, respectively. {\bf b}, Points A and B
represent the initial and the target states, respectively. The
solid curve L1 and dashed curve L2 illustrate two possible
reversal routes.}
\end{figure}

%The phase space of a Stoner particle magnetization is a
%two-dimension plane,
Magnetization $\vec{M}=\vec{m}M$ of a Stoner particle
can be conveniently described by a polar angle $\theta$
and an azimuthal angle $\phi$, shown in Fig.~\ref{fig1}a,
because its magnitude $M$ does not change with time.
The dynamics of magnetization unit direction $\vec{m}$
is governed by the dimensionless Landau-Lifshitz-Gilbert
(LLG) equation\cite{Hillebrands,xrw},
\begin{equation}
(1+\alpha^2)\frac{d\vec{m}}{dt}= -\vec{m}\times\vec{h}_{t}
-\alpha\vec{m}\times(\vec{m}\times\vec{h}_{t}),\label{LLG}
\end{equation}
where $\alpha$ is a phenomenological damping constant whose
typical value ranges from 0.01 to 0.22 for Co
films\cite{Back}. The total field $\vec{h}_{t}=\vec{h}+
\vec{h}_i$ comes from an applied field $\vec{h}$ and an internal
field $\vec{h}_i=-\nabla_{\vec{m}} w(\vec{m})$ due to the
magnetic anisotropic energy density $w(\vec{m})$.
Different particle is characterized by different $w(\vec{m})$.
In our analysis, we assume it uniaxial with the easy axis
along the z-direction, $w=w(\cos\theta)$ and $\vec{h}_i= -
\frac{\partial w(\cos\theta)}{\partial (\cos\theta)}\hat{z}
\equiv f(\cos\theta)\hat{z}$.

According to Eq.~\eqref{LLG}, each field generates
two motions, a precession motion around the field and a
damping motion toward the field as shown in Fig.~\ref{fig1}a.
In terms of $\theta-\phi$, Eq.~\eqref{LLG} can be
rewritten as\cite{Hillebrands},
\begin{align}
&(1+\alpha^2)\dot{\theta}=h_{\phi} +\alpha h_{\theta}
- \alpha f(\cos\theta) \sin\theta, \nonumber\\
&(1+\alpha^2)\sin\theta\dot{\phi}=\alpha h_{\phi}-
h_{\theta} + f(\cos\theta)\sin\theta .\label{sphe}
\end{align}
Here $h_\theta$ and $h_\phi$ are the field components
along $\hat{e}_{\theta}-$ and $\hat{e}_{\phi}-$
directions of $\vec{m}$, respectively.

The switching problem is as follows: In the absence of an
external field, the particle has two stable states,
$\vec{m}_0$ (point A) and $-\vec{m}_0$ (point B) along
its easy axis as shown in Fig.~\ref{fig1}b.
Initially, the magnetization is $\vec{m}_0$, and the goal
is to reverse it to $-\vec{m}_0$ by applying an external
field. There are infinite number of
paths that connect the initial and the target state.
L1 and L2 in Fig.~\ref{fig1}b are two examples. Each of these
paths can be used as a magnetization reversal route (path).
For a given path, there are infinite number of ways of
designing magnetic field pulses such that the magnetization
will move along the path. Different design leads to different
switching time. Let $\vec{h}^{L,s}(t)$ be the magnetic field
pulse of design $s$ along magnetization reversal route $L$.
To proceed, a few quantities must first be introduced.

{\bf\it Definition of switching field $H^{L,s}$:}
The switching field $H^{L,s}$ of design $s$ along route
$L$ is defined to be the largest magnitude of
$\vec{h}^{L,s}(t) $ for all $t$, $H^{L,s}=max\{|\vec{h}
^{L,s}(t)|; \forall t\}$.

{\bf\it Definition of minimal switching field $H^{L}$
on reversal route $L$:} The minimal switching field along
route $L$ is defined to be the smallest value of $H^{L,s}$
for all possible designs $s$ that will force the
magnetization to move along $L$, i.e.
$H^{L}=min\{H^{L,s}; \forall s\}$.

{\bf\it Definition of theoretical limit of minimal switching
field $H_c$:} The switching field limit $H_c$ is defined as
the smallest value of $H^{L}$ out of all possible routes,
i.e. $H_c=min\{H^L; \forall L\}$.

If the applied field is restricted to be static, $H^L$ forms the
so-called modified Stoner-Wohlfarth (SW) astroid\cite{field},
and $H_c$ is below, but close to, the SW field when the applied
field is about $135^{\circ}$ from the easy-axis\cite{field}.
The first issue is about the existence of $H_c$.

{\bf Theorem 1}: For a given uniaxial magnetic anisotropy
specified by $f(\cos\theta)= -\frac{\partial w(\cos\theta)}
{\partial(\cos\theta)}$, the theoretical limit of
the minimal switching field is given by $H_c =\frac
{\alpha}{\sqrt{1+\alpha^2}}Q$, where
$Q= max\{f(\cos\theta)\sin\theta \}$, $\theta \in [0,\pi]$.

{\bf Proof:} To find the lowest possible switching field, it
should be noticed that field along the radius direction $h_r$
of an external field does not appear in Eq.~\eqref{sphe}.
Thus one can lower the switching field by always putting
$h_r=0$, and the magnitude of the external field is $h=\sqrt
{h_{\theta}^2+h_{\phi}^2}$. According to Eq.~\eqref{sphe},
$\dot\theta$ and $\dot\phi$ are fully determined by
$h_{\theta}$ and $h_{\phi}$ and vice versa.
It can be shown that $h^2$ can be expressed
in terms of $\theta$, $\phi$, $\dot\theta$, and $\dot\phi$
\begin{align}
g\equiv h^2=&(1+\alpha^2)\dot{\theta}^2+2\alpha
f(\cos\theta)\sin\theta\dot{\theta}\nonumber\\
&+(\alpha\sin\theta\dot{\phi})^2
+\sin^2\theta[\dot{\phi}-f(\cos\theta)]^2.\label{h2}
\end{align}
Here $g(\dot{\theta}, \theta, \dot{\phi})$ does not
depend explicitly on $\phi$ for a uniaxial model.

In order to find the minimum of $g$, it can be shown that
$\phi$ must obey the following
equation:
\begin{equation}\label{phic}
\dot{\phi}=f(\cos\theta)/(1+\alpha^2),
\end{equation}
which is from $\frac{\partial g}{\partial\dot{\phi}}|_{(\dot
{\theta},\theta)}=0$ and
$\frac{\partial^2g}{\partial\dot{\phi}^2}|_{(\dot
{\theta},\theta)}>0$.
%It means that $g$ is the minimum if $\phi(t)$ satisfies
%Eq.~\eqref{phic} when $\theta$ and $\dot{\theta}$ are given.
%According to Eq.~\eqref{phic}, $\vec{m}$ undergoes the
%precession motion around the internal field $\vec{h}_i$.

Eq.~\eqref{phic} is a necessary condition for the smallest
minimal switching field. This can be understood as follows.
Assume $H_c$ is the minimal switching field along reversal
path $L$ described by $\theta(t)=\theta_1(t)$ and
$\phi(t)=\phi_1(t)$ (i.e. $H_c$ is the maximum magnitude of
the external field that generates the motion of $\theta_1(t)$
and $\phi_1(t)$). If $\phi_1(t)$ does not satisfy
Eq.~\eqref{phic}, then one can construct another reversal
path $L^*$ specified by $\theta(t)=\theta_1(t)$ and $\phi
(t)=\phi_2(t)$, where $\phi_2(t)$ satisfies Eq.~\eqref{phic}.
Because $\theta(t)$ and $\dot\theta$ are exactly the
same on both paths $L$ and $L^*$ at an arbitrary
time $t$, the values of $g(t)$ shall be smaller on
$L^*$ than those on $L$ at any $t$. Thus, the maximum
$g^*=(H_c^*)^2$ of $g$ on $L^*$ will be also smaller
than that ($H_c^2$) on $L$, i.e. $H_c^*<H_c$.
But this is in contradiction with the assumption that $H_c$
is the theoretical limit of the minimal switching field.
Hence, $\phi(t)$ must obey Eq.~\eqref{phic} on the optimal
path that generates the smallest switching field, $H_c$.

Substituting Eq.~\eqref{phic} into Eq.~\eqref{h2}, we have
\begin{equation}
h^2=[\sqrt{1+\alpha^2}\dot{\theta}+\alpha
f(\cos\theta)\sin\theta/\sqrt{1+\alpha^2}]^2.\label{h3}
\end{equation}
In order to make $h$ minimum, one notices that Eq.~\eqref{h3} does
not contain $\phi$. Thus optimal reversal path should satisfy
$\dot{\theta}\ge 0$ during its reversal from $\theta=0$ to
$\theta=\pi$. Furthermore, the second term in Eq.~\eqref{h3} is
anti-symmetric about $\theta=\pi/2$ plane (positive in the upper
hemisphere and negative in the lower hemisphere). Thus, the
largest value of $h$ on the optimal reversal path occurs in the
upper hemisphere. It equals $\frac{\alpha Q}{ \sqrt{1+\alpha^2}}$,
where $Q\equiv max\{f(\cos\theta) \sin\theta\}$. ---QED
%Thus, the theoretical limit of the switching field $H_c$ is
%\begin{equation}
%h_{min}=\sqrt{1+\alpha^2}0^+ +(\alpha Q)/\sqrt{1+\alpha^2}\approx
%(\alpha Q)/\sqrt{1+\alpha^2}.  \quad --Q.E.D.
%\end{equation}

\begin{figure}[htbp]
 \begin{center}
\includegraphics[width=6.5cm, height=4.5cm]{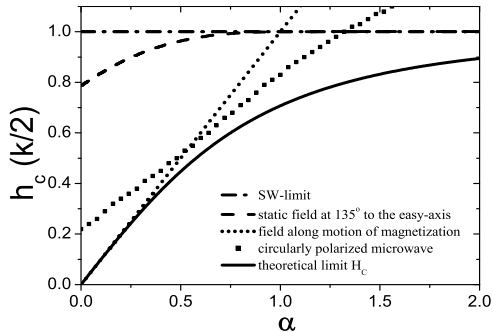}
 \end{center}
\caption{\label{fig2} The switching field $h_c$ vs. damping
constant $\alpha$ under different reversal schemes.}
\end{figure}

To have a better picture about what this theoretical limit $H_c$
is, we consider a well-studied uniaxial model, $w(\vec{m})=-k
m_z^2/2$, or $f=k \cos\theta$. It is easy to show that
the largest $h$ is at $\theta=\pi/4$ so that $Q=k/2$, and
\begin{equation}\label{limit}
H_c=\frac {\alpha}{\sqrt{1+\alpha^2}} \frac{k}{2}.
\end{equation}
At small damping, $H_c$ is proportional to the damping constant.
The result in the limit of $\alpha\rightarrow 0$ coincides with
the switching field in reference 9 where the  %\cite{xrw1}
time-dependent field always follows the motion of magnetization.
At the large damping, $H_c$ approaches the SW field\cite{xrw}
when a non-collinear static switching field is $135^{\circ}$
from the easy-axis. The solid curve in Fig.~\ref{fig2}
is $H_c$ versus $\alpha$. For comparison, the minimal
switching fields from other reversal schemes are also plotted.
The dotted line is the minimal switching field when the applied
field is always parallel to the motion of the magnetization\cite
{xrw1}. The curve in square symbols is the minimal switching
field when a circularly polarized microwave at optimal
frequencies is applied\cite{xrw1}. The dashed line is minimal
switching field under a non-collinear static field of $135^
{\circ}$ to the easy-axis. It saturates to the SW field beyond
$\alpha_c$\cite{field,xrw}.

Although the theoretical limit of the switching field is
academically important because it provides a low bound to the
switching field so that one can use the theorem to evaluate
the quality of one particular strategy, a design using a
field at the theoretical limit would not be interesting from
a practical point of view because the switching time would
be infinite long. Thus, it is more important to design a
reversal path and a field pulse such that the reversal time
is the shortest when the field magnitude H ($H>H_c$) is given.
An exact result is given by the following theorem.

{\bf Theorem 2:} Suppose a field magnitude $H$ does not
depend on time and $H> H_c$. The optimal reversal path
(connects $\theta=0$ and $\theta=\pi$) that gives the
shortest switching time is the magnetization trajectory
generated by the following field pulse $\vec{h}(t)$,
\begin{align}
&h_r(t)=0,\nonumber\\
&h_{\theta}(t)=\alpha H/\sqrt{1+\alpha^2},\label{pulse}\\
&h_{\phi}(t)=H/\sqrt{1+\alpha^2}=h_{\theta}/\alpha.\nonumber
\end{align}

{\bf Proof:} The reversal time from A to B (Fig.\ref{fig1}b)
is $T \equiv \int_0^{\pi} d\theta/\dot{\theta}.$ According to
Eq.~\eqref{sphe}, one needs $(h_{\phi}+\alpha h_{\theta})$ to
be as large as possible in order to make $\dot{\theta}$
maximal at an arbitrary $\theta$. Since $H^2=h_r^2+h_{\theta}
^2+h_{\phi}^2$, one has the following identity:
\begin{equation}
(1+\alpha^2) H^2=(1+\alpha^2)h_r^2 +(h_{\phi} +\alpha
h_{\theta})^2 +(h_{\theta}-\alpha h_{\phi})^2.
\end{equation}
Thus, $(h_{\phi}+\alpha h_{\theta})$ reaches the maximum of
$\sqrt{1+\alpha^2}H$ when $h_r=0$ and $h_{\theta}=\alpha h_
{\phi}$, which lead to Eq.~\eqref{pulse}, are satisfied. ---QED

\begin{figure}[htbp]
 \begin{center}
\includegraphics[width=6.cm, height=4.cm]{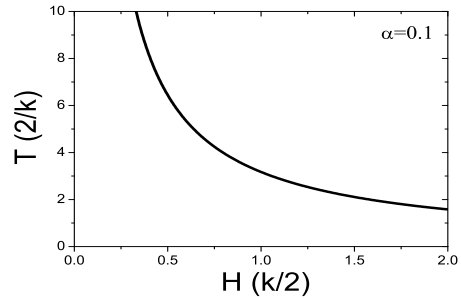}
 \end{center}
\caption{\label{fig3} The field dependence of $T$ under the
optimal field pulse Eq.~\eqref{pulse} for $\alpha=0.1$. The field
is in the unit of $k/2$ and the unit for time is $2/k$.}
\end{figure}

Under the optimal design of \eqref{pulse}, $\phi(t)$ and
$\theta(t)$ satisfy, respectively, Eq. ~\eqref{phic} and
\begin{equation}\label{dottheta} \dot{\theta} =
H/\sqrt{1+\alpha^2} -\alpha f(\cos\theta)\sin\theta/(1+\alpha^2).
\end{equation}
For uniaxial magnetic anisotropy $w(\vec{m})=-km_z^2/2$, it is
straight forward to integrate Eq.~\eqref{dottheta}, and to find
the reversal time $T$ from A to B (Fig.\ref{fig1}b),
\begin{equation}\label{time}
T=\frac{2}{k} \frac{(\alpha^2+1)\pi}{
\sqrt{4(\alpha^2+1)H^2/k^2-\alpha^2}}.
\end{equation}
In the weak damping limit $\alpha \rightarrow 0$, $T\approx\pi/
H$ while in the large damping limit $\alpha\rightarrow\infty$,
$T\approx \frac{\alpha \pi}{\sqrt{H^2-k^2/4}}\rightarrow\infty$.
For the large field $H\rightarrow\infty$, $T\approx \frac{\sqrt{
\alpha^2+1}\pi}{H}$, inversely proportional to the field strength.
Thus, it is better to make $\alpha$ as small as possible.
Then the critical field is low, and the speed is fast ($T\sim
\pi/H$). Fig.~\ref{fig3} shows the field dependence of the
switching time for $\alpha=0.1$, where $T$ and $H$ are in
the units of $2/k$ and $k/2$, respectively.

How much could the so-called ballistic (precessional) reversal
strategy\cite{field,Schumacher}
be improved? To answer the question, let us compare the switching
field and time in the ballistic reversal with those of theoretical
limits for uniaxial magnetic anisotropy $w(\vec{m})=-km_z^2/2$ and
$\alpha=0.1$. According to reference 8, the smallest %\cite{xrw}
switching field (in unit of $k/2$) for the ballistic connection
between A and B (Fig.\ref{fig1}b) is $H=1.02$ applied in
$97.7^{\circ}$ to the easy z-axis, and the corresponding ballistic
reversal time (in unit of $2/k$) is $T=5.87$. On the other hand,
the theoretical limit for the minimal switching field is
$H_c\approx 0.1$ from Eq.~\eqref{limit}, about one tenth of the
minimal switching field in the ballistic reversal\cite{xrw}. For
realistic value of $\alpha$ of order of $0.01$, the difference
between experimentally achieved low switching field and the
theoretical limit is of the order of hundred times. Thus there is a
very large room for an improvement. It is also possible to switch
a magnetization faster than that of the conventional ballistic
reversal by using a smaller field. For example, To achieve a
reversal time of $T=5.87$ along the optimal route, the field
magnitude can be as lower as $H=0.547$ (instead of $H=1.02$)
according to Fig.~\ref{fig3}.
%This is about half of that in the ballistic reversal.

\begin{figure}[htbp]
 \begin{center}
\includegraphics[width=6.cm, height=4.cm]{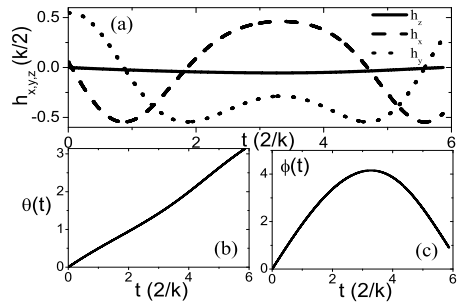}
 \end{center}
\caption{\label{fig4} Time dependence of different field
components,  $\theta$ and $\phi$ for uniaxial magnetic anisotropy
$w(\vec{m})=-km_z^2/2$ with $\alpha=0.1$ and $H=0.547$ when the
reversal path is optimal. The reversal time is $T=5.87$. {\bf a},
X-, y- and z-components of magnetic field. {\bf b}, $\theta(t)$.
{\bf c}, $\phi(t)$. }
\end{figure}

The field pulse given in Eq.~\eqref{pulse} requires a constant
adjustment of field-direction during the magnetization reversal.
To have a better ideal about the type of fields required, we plot
in Fig. \ref{fig4}a the time dependence of x-, y- and z-component
of the field while its magnitude is kept at $H=0.547$. The time
dependence of $\theta$ and $\phi$ is also plotted in
Fig.~\ref{fig4}b-c.

%{\it {Discussion and conclusions--}}
Although the Stoner-Wohlfarth problem of magnetization reversal
for a uniaxial model is of great relevance to the magnetic
nano-particles, it is interesting to generalize the results to the
non-uniaxial cases. So far, our results is on the magnetic field
induced magnetization reversal, it will be extremely important to
generalize the results to the spin-torque induced magnetization
reversal. It should also be pointed out that it is an experimental
challenge to create a time-dependent field pulse given by
Eq.~\eqref{pulse} in order to implement the optimal design
reported here.

In conclusion, the theoretical limit of the magnetization
switching field for uniaxial Stoner particles is obtained. The
limit is proportional to the damping constant at weak damping and
approaches the SW field at large damping. When the field magnitude
is kept to a constant, and the field direction is allow to vary,
the optimal field pulse and reversal time are obtained.

{\it{Acknowledgments}--}This work is supported by UGC, Hong Kong,
through RGC CERG grants. Discussion with Prof. J. Shi is
acknowledged.

\end{document}